\documentclass[referee]{aa}  

\usepackage{graphicx}

\usepackage{txfonts}

\begin{document}

\headnote{Research Note}

\title{The mid-UV population of the nucleus and the bulk of the post-merger NGC~3610}

   \author{L.M.~Buson
   }

   \offprints{Lucio M. Buson}

   \institute{
   INAF Osservatorio Astronomico di Padova, vicolo
   dell'Osservatorio~5, I-35122 Padova, Italy\\
   \email{lucio.buson@oapd.inaf.it}
   }
   \date{Received ... ; accepted .... 2010}

   \titlerunning{UV Spectrum of NGC 3610}
   \authorrunning{L. M. Buson}

  \abstract
    {The very center (r$\ll$r$_e$) of NGC~3610, a clearly disturbed giant elliptical generally assumed to be a post-merger
     remnant, appears dominated in the mid-UV (2500-3200~\AA\ spectral region) by a stellar population markedly
     different from that dominating the bulk of its stellar body.}
    {I want to make use of  the mid-UV spectra  of NGC~3610 as seen through tiny ($\sim$1") and large (10"$\times$20") apertures as a diagnostic population tool.}
    {I compare archive IUE/LWP large aperture and HST/FOS UV data of NGC 3610.}
    {The strength of mid-UV triplet (dominated by the turnoff population) shows a remarkable drop  
    when switching  from the galaxy central arcsec (FOS aperture) 
    to an aperture size comparable to $\sim$0.5 r$_e$ (IUE).}
    {The sub-arsec (mid)-UV properties of this galaxy involved in a past merger reveal
      a central metal enrichment which left intact the bulk of its pre-existing population.}  
  

   \maketitle

\section{Introduction}

\noindent
The E5 galaxy NGC~3610 appears peculiar in many respects. It has the 
richest fine structure of all ellipticals discussed by Seitzer \& 
Schweizer (1990), as well as a warped central disk identified by Scorza 
\& Bender (1990) by means of their photometric decomposition and later 
confirmed by means of HST by Whitmore {\em et al.} (1997).  In addition,  its 
anomalous (B-V) color (e.g.  Goudfrooij et al. 1994) is an indicator of recent
star formation. At the same time the galaxy is encircled by a globular cluster (GC) system
showing evidence of a past merger of metal-rich disk-disk galaxies (Goudfrooij et al. 2007).\\

Since the epoch of the major merger which formed NGC~3610 is estimated 
to be 4 $\pm$ 2.5 Gyr ago (Strader et al. 2004), this makes it an ideal galaxy for studying 
this kind of phenomenon, including the so-called ``formation by merger''
scenario for metal-rich GCs (Schweizer 1987; Ashman \& Zepf 1992). Such galaxies,  though having reached a substantial
dynamical equilibrium, are indeed still young enough to keep track, both
morphological and evolutionary, of the past interaction.\\ 

In fact, though the most exhaustive, up-to-date study 
of the age, metallicity and $\alpha$-enhancement gradients within NGC~3610 of Howell et al. (2004) does not find 
significant stellar population gradients in the outer parts of the galaxy (r~$\geq$~0.75 r$_e$),
a clear indication of the presence of a {\em central} young stellar population in NGC~3610 has been pointed out by Silva 
\& Bothun (1998) on the basis of the strong central H$\beta$ absorption in the optical and the IR excess 
$\Delta$(H-K) between (r~$\leq$ 0.5 kpc = 0.25 r$_e$) and the outer annulus (1.0 $\leq$ r $\leq$ 1.5 kpc)
induced by the presence of intermediate-age AGB stars.
Silva \& Bothun (1998) refer also to a central, subarcsec UV HST/FOS spectrum of the galaxy they put forward as a proof 
of intermediate-age nuclear stars, being well-matched by a mid-F main-sequence 
stellar population very similar to that of M32.\\

Luckily enough the existing (archive), space-borne ultraviolet data include also a good-quality 
UV spectrum of the galaxy through the large (10$\times$20'') IUE/LWP aperture giving the 
luminosity-averaged spectral energy distribution (SED) within an aperture photometrically equivalent to an
aperture 7'' in radius (cf. Burstein et al. 1988).
Since the two UV spectra, especially as far as the region of the
absorption lines formed by the Mg~II 2800~\AA, Mg~I 2852~\AA\ and the Fe~I 
$+$~II $+$~Cr~I feature at 2750~\AA~(the so-called UV triplet, well-known for showing rapid changes 
in relative strengths from late B to late F stars) is concerned, appear
quite different (see Fig.~1), one can expect their comparison represents a further proof of 
the different nature of the galaxy's inner and large-aperture stellar population.  \\

\begin{figure*}
   \centering
   \includegraphics[width=14cm,height=14cm]{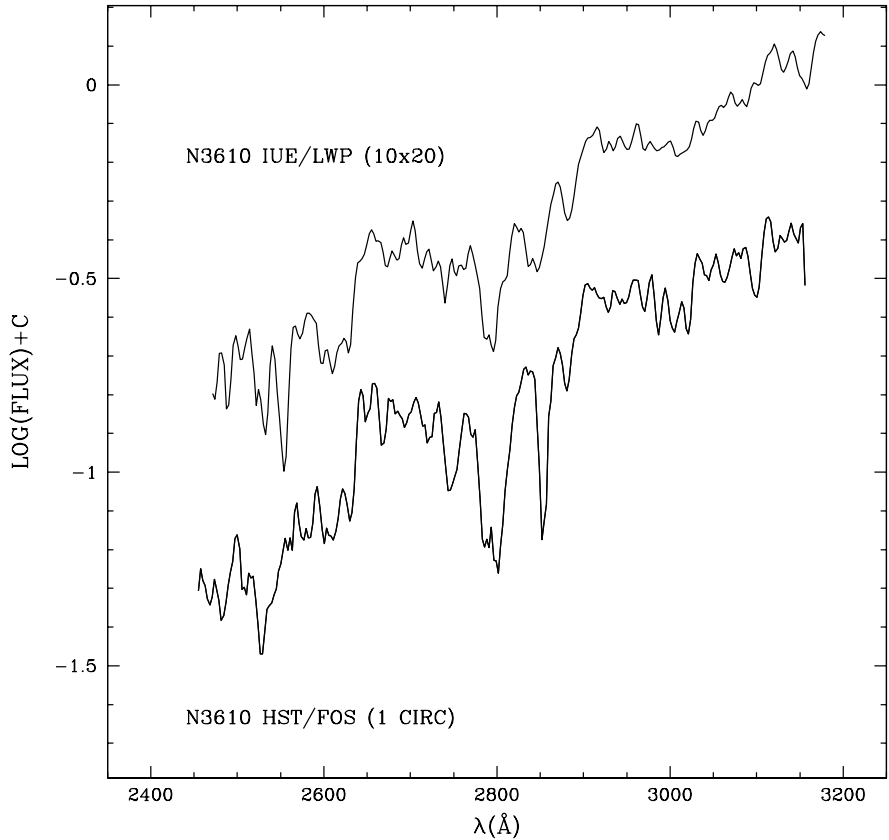}
   \caption{Comparison of the IUE /LWP (10$''$$\times$20$''$) spectrum
   of NGC~3610 and its $\sim$1$''$ circular central HST/FOS~G270H spectrum. 
   Note the outstanding difference of the mid-UV triplet between the two 
   spectra of the same object. All plots are given in log F$_\lambda$$+$C.
   The HST/FOS G270H spectrum has been properly smoothed to match 
   the IUE/LWP spectral resolution (about 6-7~\AA\ about $\lambda$ 2,700~\AA).
   An arbitrary shift has been finally applied for showing purposes.}
\label{fig:iue_fos}
\end{figure*}

\section{UV observations}

A good-quality (15,300~s exposure time) IUE/LWP spectrum of NGC~3610 has been 
obtained in 1992 at GSFC by LB. The orientation of the IUE large 
aperture was P.A.=154$^{\circ}$. As stressed above its $''$oval$''$ (10$''$$\times$20$''$) shape 
is photometrically equivalent to a 7$''$ radius circular aperture ({\em cf.} 
Burstein {\em et al.} 1988), {\em i.e.} $\sim$0.5~r$_e$ in the case of NGC~3610 
(Idiart {\em et al.} 2003). As in the case of NGC~5018 (Bertola {\em et al.} 1993), a
IUE/SWP spectrum shortward of $\lambda~$1800~\AA\ provided no signal, 
thus implying that also NGC~3610 lacks the prominent UV-upturn typical of old, 
metal-rich giant spheroids. \\\

Almost simultaneously NGC~3610 was observed with 
the HST/FOS on 1992 November 30. A FOS circular aperture 0$''$.86 diameter has been used. The target acquisition procedure involved measuring the flux on a 4$\times$4 grid 
at 0$''$.25 intervals so as to center the galaxy to an accuracy of 0$''$.25. The datum 
I use here is the mid-UV FOS/G270H spectrum. \\

The properly extracted, redshift-corrected IUE and HST/FOS mid-UV spectra of NGC~3610 are shown on a logarithmic scale in Fig.~1 (they are properly shifted for showing purposes) after a proper resolution match of the original data. 
Following both Burstein \& Heiles (1984) 
and Schlegel et al. (1998),  NGC~3610 is galactic extinction-free and no correction has been applied.
A quantitative comparison of commonly used mid-UV indices is instead  given in Tab.~1 for both apertures. 
The same set of indices is given also for the center of M32 for comparison with an intermediate-age population.\\

NGC 3610 has not been observed by the GALEX satellite yet, thus hampering the possibility of a useful comparison of its optical metallicity gradient with an extended ultraviolet color (FUV$-$NUV) profile.\\

\begin{table}[h!]
\caption{IUE vs. FOS Mid-UV Indices}
\begin{tabular}{lccc}
\hline
\noalign{\smallskip}
\multicolumn{1}{c}{Feature} &
\multicolumn{1}{c}{N3610/IUE} &
\multicolumn{1}{c}{N3610/FOS}&
\multicolumn{1}{c}{M32/FOS} \\
\multicolumn{1}{c} {} & 
\multicolumn{1}{c} {mag} & 
\multicolumn{1}{c} {mag} & 
\multicolumn{1}{c} {mag}\\ 
\noalign{\smallskip}
\hline
\hline
\noalign{\smallskip}
MgWide & 0.351$\pm$0.017 & 0.370$\pm$0.007 & 0.285$\pm$0.002\\
FeI+II+CrI & 0.082$\pm$0.050 & 0.188$\pm$0.018 & 0.351$\pm$0.010\\
MgII2800   & 0.442$\pm$0.047 & 0.756$\pm$0.018 & 0.795$\pm$0.014\\
MgI2852    & 0.237$\pm$0.046 & 0.456$\pm$0.017 & 0.349$\pm$0.008\\
FeI3000 & 0.106$\pm$0.040 & 0.208$\pm$0.013 & 0.199$\pm$0.003\\
2828/2921 & 0.561$\pm$0.028 & 0.577$\pm$0.011 & 0.510$\pm$0.005\\
\noalign{\smallskip}
\hline
\noalign{\smallskip}
\end{tabular}\\
\label{tab:fos_indc}
\end{table}

\section{Mid-UV indices as a population diagnostic tool}

From Fig~1 it is immediately recognizable that the FOS (central) spectrum shows a much deeper mid-UV triplet
close to $\lambda$~2800~\AA\ which reflects in the indices of Tab.~1 computed following the definition of Chavez et al. (2007).
As far as the index measurements are concerned, it is well known that IUE cameras (and thus extracted spectra) 
are potentially affected by the camera reseau marks. Luckily enough I verified that the LWP camera artifacts do not affect
the mid-UV indices I made use of in this paper (from Fig.~2 one can see that neither the red bandpass of the index 
FeI3000 (ending at $\lambda$ 3051~\AA) turns out to be influenced).

A careful analysis of the indices listed in Table~1 does suggest the following picture for the chemistry/age pattern
of the center and bulk of NGC~3610, as well as for the comparison object M32: (i) on the basis of the doubling of the main iron and magnesium indices like FeI+II+CrI, FeI3000, MgII2800 and MgI2852 the central regions (within the FOS aperture)
of NGC~3610 appear systematically more metal rich (when compared with the large IUE region) and---{\bf assuming that the 
metallicity of the (central) population is approximately solar (cf. Fig~4 in Howell et al. 2004)}---younger than 4~Gyr 
(cf. Fig.~2 of Chavez et al. 2009); (ii) The nucleus of NGC~3610 appears quite $\alpha$-enhanced in comparison with that of M32 taking into account, e.g., the much lower value of the iron index FeI+II+CrI and the slight increase of the MgWide index
(cf. Fig.~5 of Chavez et al. 2009).

\begin{figure}
   \centering
   \includegraphics[width=8.5cm,height=8.5cm]{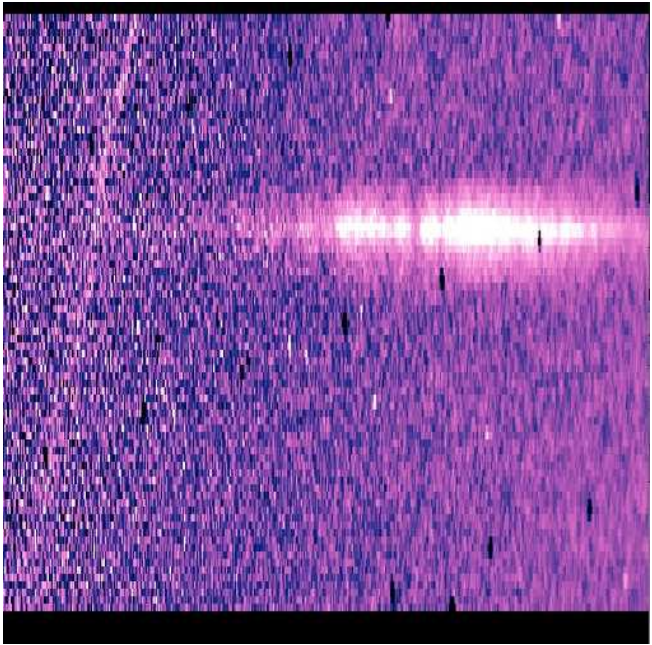}
   \caption{IUE /LWP10105 spectrum of M32 shown here for showing
   purposes (i.e. to give the
   right localization of the camera reseau marks). As can be seen the
   spectral regions corresponding to the indices used here are free from
   reseau contamination.}
\label{fig:iue_fos2}
\end{figure}

\section{Comparison with NGC 5018}
A similar analysis, namely IUE Large Aperture vs. FOS,  has been adopted by Buson et al. (2004) for NGC~5018, an object with many similarities with NGC~3610 (e.g. a major merger suffered $\sim$3~Gyr ago [Leonardi \& Worthey 2000]). 
However, unlike NGC~3610, the stellar population of NGC~5018 tells a simple story, in the sense that it shows both a strict match of its IUE and HST/FOS circumnuclear spectrum and quite a shallow metallicity 
gradient, as traced by Mg$_2$ index (dMg$_2$/dlogr=-0.04; Carollo \& Danziger 1994).  In other words, its 
population is very homogeneous, both in age and metal content (as expected to occur if the merger
is largely dissipationless) and/or suffered from a rejuvenation involving the whole galaxy.\\

In this respect NGC~3610 could instead represent a different  kind  of  merger, especially taking into account
that its inner mean gradient decreases at a rate of [Z/H]=-0.30 per decade in r/r$_e$ (Howell et al. 2004).
The onset of such a gradient, though not unexpected during a dissipative galaxy merger (e,g.  Barnes \& Hernquist 1991;
Mihos \& Hernquist 1994; Kobayashi 2004), represents unequivocally a population unhomogeneity. 
Moreover, its stronger absorption mid-UV spectrum appears confined to its very  center alone.
{\bf The natural conclusion is that the central arcsec in NGC~3610 underwent significantly more 
rejuvenation and/or metal enrichment than its surroundings, unlike NGC~5018.}

\begin{acknowledgements}It is a pleasure to thank Sandro Bressan---largely experienced in mid-UV indices modeling---for useful discussions. I thank the anonymous referee for his/her comments improving this paper.

\end{acknowledgements}

\end{document}